\RequirePackage{ifpdf}
\ifpdf 
\documentclass[pdftex]{sigma}
\else
\documentclass{sigma}
\fi

\begin{document}
\allowdisplaybreaks


\def\eqrefs#1#2{(\ref{#1}) and~(\ref{#2})}
\def\eqsref#1#2{(\ref{#1}) to~(\ref{#2})}
\def\sysref#1#2{(\ref{#1})--(\ref{#2})}

\def\Ref#1{Ref.~\cite{#1}}

\hyphenation{Eq Eqs Sec App Ref Fig}

\def\downupindices#1#2{{\mathstrut}^{}_{#1}{\mathstrut}_{}^{#2}}
\def\updownindices#1#2{{\mathstrut}_{}^{#1}{\mathstrut}^{}_{#2}}
\def\mixedindices#1#2{{\mathstrut}^{#1}_{#2}}
\def\downindex#1{{\mathstrut}_{#1}}
\def\upindex#1{{\mathstrut}^{#1}}

\def\eqtext#1{\hbox{\rm{#1}}}

\def\hp#1{\hphantom{#1}}

\def\frame#1#2{e\mixedindices{#2}{#1}}
\def\xframe#1{e_x\upindex{#1}}
\def\tframe#1{e_t\upindex{#1}}
\def\coframe#1{e\downindex{#1}}
\def\conx#1#2{\omega\downupindices{#1}{#2}}
\def\xconx#1#2{\omega_x\downupindices{#1}{#2}}
\def\tconx#1#2{\omega_t\downupindices{#1}{#2}}
\def\gcovder#1{{}^g\nabla\downindex{#1}}
\def\covder#1{\nabla\downindex{#1}}
\def\v#1#2{v\mixedindices{#1}{#2}}
\def\bdsymb#1{{\boldsymbol{#1}}}
\def\curv#1#2{R\downupindices{#1}{#2}}
\def\w#1#2{\varpi\mixedindices{#1}{#2}}
\def\vs#1{{\mathfrak {#1}}}
\def\gconx{{}^{\vs{g}}\omega}
\def\D#1{D\downindex{#1}}
\def\Dinv#1{D\downindex{#1}^{-1}}
\def\trans#1{{#1}{}^T}
\def\vvec{\vec{v}}
\def\hvec{\vec{h}}
\def\wvec{\vec{\varpi}}
\def\Hop{{\mathcal H}}
\def\Jop{{\mathcal J}}
\def\Rop{{\mathcal R}}
\def\Eop{{\mathcal E}}
\def\Rnum#1{{\mathbb R}\upindex{#1}}
\def\idmatr{{\mathbb I}}
\def\h#1#2{h\mixedindices{#1}{#2}}
\def\hpar{h_\parallel}
\def\hperp{h_\perp}
\def\tr{{\rm tr}}
\def\ad{{\rm ad}}
\def\covD#1{{\mathcal D}\downindex{#1}}
\def\hook{\lrcorner}
\def\i{{\rm i}}

\renewcommand{\PaperNumber}{044}

\FirstPageHeading

\ShortArticleName{Hamiltonian Flows and Vector Soliton Equations}

\ArticleName{Hamiltonian Flows of Curves in symmetric spaces\\
$\boldsymbol{G/SO(N)}$ and Vector Soliton Equations of mKdV\\
and Sine-Gordon Type}

\Author{Stephen C. ANCO}
\AuthorNameForHeading{S.C. Anco}

\Address{Department of Mathematics, Brock University, Canada}
\Email{
{sanco@brocku.ca}}

\ArticleDates{Received December 12, 2005, in f\/inal form April
12, 2006; Published online April 19, 2006}

\Abstract{The bi-Hamiltonian structure of the two known vector generalizations
of the mKdV hierarchy of soliton equations is derived in a
geometrical fashion from f\/lows of non-stretching curves in
Riemannian symmetric spaces $G/SO(N)$.
These spaces are exhausted by the Lie groups  $G=SO(N+1),SU(N)$.
The derivation of the bi-Hamiltonian structure uses
a parallel frame and connection along the curves,
tied to a zero curvature Maurer--Cartan form on $G$,
and this yields the vector mKdV recursion operators 
in a geometric $O(N-1)$-invariant form.
The kernel of these recursion operators is shown to yield two
hyperbolic vector generalizations of the sine-Gordon equation.
The corresponding geometric curve f\/lows in the hierarchies are described
in an explicit form, given by wave map equations and mKdV analogs
of Schr\"odinger map equations.}

\Classification{37K05; 37K10; 37K25; 35Q53; 53C35}

\vspace{-2mm}

\section{Introduction}

There has been much recent interest in the close relation between integrable
partial dif\/ferential equations and the dif\/ferential geometry of
plane and space curves
(see \cite{ChouQu1,ChouQu2,ChouQu3,ChouQu4,SandersWang1}
for an overview and many results).
The present paper studies f\/lows of curves in Riemannian manifolds 
$G/SO(N)$ for arbitrary $N \geq 2$, 
where $G=SO(N+1),SU(N)$. 
Such symmetric spaces \cite{Helgason} are well-known to exhaust
all examples of curved $G$-invariant geometries 
that are a natural generalization of Euclidean spaces
$\Rnum{N} \simeq Euc(N)/SO(N)$
modeled by replacing the Euclidean isometry group 
with a compact semisimp\-le Lie-group $G \supset SO(N)$.

It will be shown that if non-stretching curves are described using 
a moving parallel frame and an associated frame connection $1$-form 
in $G/SO(N)$ then the frame structure equations for torsion and
curvature encode $O(N-1)$-invariant bi-Hamiltonian operators.
These operators will be demonstrated to produce a hierarchy of
integrable f\/lows of curves in which the frame components of the
principal normal along the curve satisfy $O(N-1)$-invariant vector
soliton equations.
The hierarchies for both $SO(N+1)/SO(N)$, $SU(N)/SO(N)$ will be seen
to possess a scaling symmetry and accordingly will be organized by
the scaling weight of the f\/lows.
The $0$~f\/low just consists of a convective (traveling wave)
equation, while the $+1$ f\/low will be shown to give the two vector
generalizations of the mKdV equation known from
symmetry-integrability classif\/ications of vector evolution equations
in \cite{SokolovWolf}.
A recent classif\/ication analysis \cite{AncoWolf} found there are
vector hyperbolic equations for which the respective
vector mKdV equations are higher symmetries.
These two vector hyperbolic equations will be shown to describe
a $-1$ f\/low in the respective hierarchies for $SO(N+1)/SO(N)$ and
$SU(N)/SO(N)$.

As further results, the Hamiltonian operators will yield explicit
$O(N-1)$-invariant recursion operators for higher symmetries and
higher conservation laws of the vector mKdV equations and the vector
hyperbolic equations.  The associated curve f\/lows produced from
these equations will describe geometric nonlinear PDEs, in
particular given by wave maps and mKdV analogs of Schr\"odinger maps.

Previous fundamental work on 
vector generalizations of KdV and mKdV equations 
as well as their Hamiltonian structures and geometric origin 
appeared in \cite{AthorneFordy,Athorne,SandersWang2,SandersWang3}.
In addition, the bi-Hamiltonian structure of both vector mKdV equations
was f\/irst written down in \cite{Wang} 
from a~more algebraic point of view,
in a multi-component (non-invariant) notation. 
Special cases of two component KdV--mKdV integrable systems
related to vector mKdV equations have been discussed recently in
\cite{Foursov,TsuchidaWolf,SergyeyevDemskoi}.

\section[Curve flows, parallel frames, and Riemannian symmetric
spaces]{Curve f\/lows, parallel frames,\\ and Riemannian symmetric
spaces}

Let $\gamma(t,x)$ be a f\/low of a non-stretching curve in some
$n$-dimensional Riemannian manifold $(M,g)$.
Write $Y=\gamma_{t}$ for the evolution vector of the curve and write
$X=\gamma_{x}$ for the tangent vector along the curve normalized by
$g(X,X)=1$, which is the condition that $\gamma$ is non-stretching,
so thus $x$ represents arclength.
In the tangent space $T_\gamma M$ of the two-dimensional surface
swept out by $\gamma(t,x)$ we introduce 
orthonormal frame vectors $\frame{a}{}$ 
and connection $1$-forms $\conx{}{ab}=\conx{}{[ab]}$ 
related through the Riemannian covariant
derivative operator $\gcovder{}$ in the standard way 
\cite{KobayashiNomizu}:
\begin{gather*} 
\gcovder{x} \frame{a}{} = (X \hook \conx{a}{b}) \frame{b}{} ,\qquad
\gcovder{t} \frame{a}{} = (Y \hook \conx{a}{b}) \frame{b}{} .
\end{gather*}
(Throughout, $a,b=1,\ldots,n$ denote frame indices which get
raised and lowered by the Euclidean metric
$\delta\downindex{ab}={\rm diag}(+1,\ldots,+1)$).
Now choose the frame along the curve to be parallel \cite{Bishop}, so it
is adapted to $\gamma$ via 
\begin{gather*}
\frame{a}{} := X\ \ (a=1),\qquad (\frame{a}{})_{\perp}\ \ (a=2,\ldots,n)
\end{gather*}
where $g(X,(\frame{a}{})_{\perp})=0$, such that the covariant
derivative of each of the $n-1$ normal vectors~$(\frame{a}{})_\perp$
in the frame is tangent to $\gamma$, 
\begin{gather}\label{parallelconxtang}
\gcovder{x} (\frame{a}{})_{\perp} =-\v{}{a} X 
\end{gather}
holding for some functions $\v{}{a}$,
while the covariant derivative of the tangent vector
$X$ in the frame is normal to $\gamma$, 
\begin{gather}\label{parallelconxperp}
\gcovder{x} X =
\v{a}{}(\frame{a}{})_{\perp} .
\end{gather}
Equivalently, along $\gamma$
the connection $1$-forms of the parallel frame 
are given by the skew matrix 
$\xconx{}{ab}:=X \hook \conx{}{ab} = 2 \xframe{[a} \v{b]}{}$
where $\xframe{a}:=g(X,\frame{}{a})$ is the row matrix of the frame
in the tangent direction.
In matrix notation we have
\begin{gather}\label{riemannframe}
\xframe{a}=(1, \vec{0}) ,\qquad
\xconx{a}{b}=
\begin{pmatrix}
0 & \v{b}{}\\ -\v{}{a}&\bdsymb{0} 
\end{pmatrix} ,
\end{gather}
with $\vec{0}$, $\bdsymb{0}$ respectively denoting 
the $1 \times (n-1)$ zero row-matrix and $(n-1) \times (n-1)$ zero skew-matrix.
(Hereafter, upper/lower frame indices will represent 
row/column matrices.)
This matrix description \eqref{riemannframe} 
of a parallel frame 
has a purely algebraic characterization:
$\xframe{a}$ is a f\/ixed unit vector in $\Rnum{n}$ 
preserved by a $SO(n-1)$ rotation subgroup of 
the local frame structure group $SO(n)$,
while $\xconx{a}{b}$ belongs to the orthogonal complement of 
the corresponding rotation subalgebra $\vs{so}(n-1)$ 
in the Lie algebra $\vs{so}(n)$ of $SO(n)$.

The curve f\/low has associated to it 
the pullback of the Cartan structure equations \cite{KobayashiNomizu}
expressing that the covariant derivatives $\gcovder{x}:=X \hook
\gcovder{}$ along the curve and $\gcovder{t}:=Y \hook \gcovder{}$
along the f\/low have vanishing torsion
\begin{gather}\label{torseq} 
\gcovder{x} \gamma_{t} - \gcovder{t} \gamma_{x} = [X,Y] = 0
\end{gather}
and carry curvature determined from the metric $g$, 
\begin{gather}\label{curveq} 
[\gcovder{x},\gcovder{t}]=\curv{}{}(X,Y) 
\end{gather}
given by the Riemann tensor $\curv{}{}(X,Y)$
which is a linear map on $T_x M$ depending bilinearly on~$X$,~$Y$.  
In frame components the torsion and curvature equations look like
\cite{KobayashiNomizu} 
\begin{gather}\label{frametorseq} 
0 = \D{x} \tframe{a} - \D{t} \xframe{a} + \tframe{b}
\xconx{b}{a} - \xframe{b} \tconx{b}{a} ,
\\
\label{framecurveq}
\curv{a}{b}(X,Y) = 
\D{t} \xconx{a}{b} - \D{x} \tconx{a}{b} 
+\tconx{a}{c} \xconx{c}{b} - \xconx{a}{c} \tconx{c}{b} .
\end{gather}
Here 
$\tframe{a}:= g(Y,\frame{}{a})$ and 
$\tconx{a}{b}:= Y \hook\conx{a}{b} 
= g(\frame{}{b},\gcovder{t} \frame{a}{})$ are respectively
the frame row-matrix and connection skew-matrix in the f\/low direction, 
and
$\curv{a}{b}(X,Y):= g(\frame{}{b},
[\gcovder{x},\gcovder{t}]\frame{a}{})$ is the curvature matrix.

As outlined in \cite{Anco,SandersWang2}, 
these frame equations \eqrefs{frametorseq}{framecurveq}
directly encode a bi-Hamiltonian structure based on geometrical variables
when the geometry of $M$ is characterized by having its frame
curvature matrix $\curv{a}{b}(\frame{c}{},\frame{d}{})$ be constant on $M$.
In this situation the Hamiltonian variable is given by the
principal normal 
$\v{}{}:=\gcovder{x} X = \v{a}{}(\frame{a}{})_{\perp}$ 
in the tangent direction of $\gamma$, 
while the principal normal in the f\/low direction 
$\w{}{}:= \gcovder{t} X =\w{a}{}(\frame{a}{})_{\perp}$ 
represents a Hamiltonian covector f\/ield,
and the normal part of the f\/low vector
$\hperp:=Y_{\perp}=\h{a}{}(\frame{a}{})_{\perp}$ 
represents a Hamiltonian vector f\/ield\footnote{See \cite{Dorfman,Olver} and the appendix of \cite{Anco}
for a summary of Hamiltonian theory relevant to PDE systems.}.
In a parallel frame these variables $\v{a}{}$, $\w{a}{}$, $\h{a}{}$ are
encoded respectively in the top row of the connection matrices
$\xconx{a}{b}$, $\tconx{a}{b}$, and in the row matrix
$(\tframe{a})_{\perp}= \tframe{a} - \hpar \xframe{a}$
where $\hpar:= g(Y,X)$ is the tangential part of the
f\/low vector.

A wide class of Riemannian manifolds $(M,g)$ in which the frame
curvature matrix $\curv{a}{b}(\frame{c}{},\frame{d}{})$ is constant on
$M$ consists of the symmetric spaces $M=G/H$ for compact semisimple
Lie groups $G \supset H$
(such that $H$ is invariant under an involutive automorphism of $G$).
In such spaces the Riemannian curvature
tensor and the metric tensor are covariantly constant and
$G$-invariant \cite{KobayashiNomizu}, which implies constancy of the
curvature matrix $\curv{a}{b}(\frame{c}{},\frame{d}{})$.
The metric tensor $g$ on $M$ is given by the Cartan--Killing inner
product $\langle \cdot,\cdot\rangle$ on $T_x G \simeq \vs{g}$ restricted to the Lie
algebra quotient space $\vs{p} = \vs{g}/\vs{h}$ with 
$T_x H \simeq\vs{h}$, where $\vs{g} = \vs{h} \oplus \vs{p}$ decomposes
such that $[\vs{h},\vs{p}] \subseteq \vs{p}$ 
and $[\vs{p},\vs{p}] \subseteq \vs{h}$ 
(corresponding to the eigenspaces of the adjoint action of the involutive
automorphism of $G$ that leaves $H$ invariant).
A complete classif\/ication of symmetric spaces is given in \cite{Helgason}; 
their geometric properties are summarized in \cite{KobayashiNomizu}.  
In these spaces 
$H$ acts as a gauge group so consequently the bi-Hamiltonian structure 
encoded in the frame equations 
will be invariant under the subgroup of $H$ that leaves $X$ 
f\/ixed\footnote{More details will be given elsewhere \cite{forthcoming}.}.

Thus in order to obtain $O(N-1)$-invariant bi-Hamiltonian operators,
as sought here, we need the group $O(N-1)$ to be 
the isotropy subgroup in $H$ leaving $X$ f\/ixed.
Hence we restrict attention to the symmetric spaces
$M=G/SO(N)$ with $H=SO(N) \supset O(N-1)$.
From the classif\/ication in \cite{Helgason} all examples of these spaces
are exhausted by $G=SO(N+1),SU(N)$.
The example $M=SO(N+1)/SO(N) \simeq S^{N}$ is isometric to the
$N$-sphere, which has constant curvature.
In this symmetric space, the encoding of bi-Hamiltonian operators in
terms of geometric variables has been worked out in \cite{Anco} 
using the just intrinsic Riemannian geometry of the $N$-sphere, 
following closely the ideas in \cite{SandersWang1,SandersWang2}.
An extrinsic approach based on Klein geometry \cite{Sharpe,AncoWolf}
will be used here, as it applicable to both symmetric spaces 
$SO(N+1)/SO(N)$ and $SU(N)/SO(N)$.

In a Klein geometry 
the left-invariant $\vs{g}$-valued Maurer--Cartan form on the
Lie group $G$ is identif\/ied with a zero-curvature connection
$1$-form $\omega_G$ called the Cartan connection \cite{Sharpe}.
Thus 
\begin{gather*}
0 = d\omega_G + \frac{1}{2} [\omega_G,\omega_G],
\end{gather*}
where $d$ is the total exterior derivative on the group manifold $G$.
Through the Lie algebra decomposition 
$\vs{g}=\vs{so}(N) \oplus\vs{p}$ with 
$[\vs{p},\vs{p}] \subset \vs{so}(N)$ 
and
$[\vs{so}(N),\vs{p}] \subset \vs{p}$,
the Cartan connection determines a Riemannian structure 
on the quotient space
$M = G/SO(N)$
where $G$ is regarded~\cite{Sharpe} as a principal $SO(N)$ bundle over $M$.
Fix any local section of this bundle 
and pull-back $\omega_G$ to give a $\vs{g}$-valued $1$-form $\gconx{}$
at $x$ in $M$. 
The ef\/fect of changing the local section is to induce 
a~$SO(N)$ gauge transformation on $\gconx{}$.
Let $\sigma$ denote an involutive automorphism of $\vs{g}$ such that
$\vs{so}(N)$ is the eigenspace $\sigma=+1$, $\vs{p}$ is the
eigenspace $\sigma=-1$.
We consider the corresponding decomposition of $\gconx{}$:  
it can be shown that \cite{Sharpe} the symmetric part
\begin{gather}\label{kleinconx} 
\conx{}{}:= \frac{1}{2}(\gconx{} + \sigma(\gconx{})) 
\end{gather}
def\/ines a $\vs{so}(N)$-valued connection $1$-form for the group
action of $SO(N)$ on the tangent space $T_x M \simeq \vs{p}$, 
while the antisymmetric part
\begin{gather}\label{kleinframe}
\coframe{}:=\frac{1}{2}(\gconx{} - \sigma(\gconx{})) 
\end{gather}
def\/ines a $\vs{p}$-valued coframe for the Cartan--Killing
inner product $\langle\cdot,\cdot\rangle_{\vs{p}}$ on $T_x G \simeq \vs{g}$
restricted to $T_x M \simeq \vs{p}$.
This inner product $\langle\cdot,\cdot\rangle_\vs{p}$ 
provides a Riemannian metric 
\begin{gather*}
g=\langle \coframe{} \otimes \coframe{}\rangle_\vs{p}
\end{gather*}
on $M=G/SO(N)$,
such that the squared norm of any vector $X\in T_x M$ is
$|X|_g^2 = g(X,X)=\langle X\hook \coframe{},X\hook \coframe{}\rangle_\vs{p}$.

Moreover there is a $G$-invariant covariant derivative $\covder{}$
associated to this structure 
whose restriction to the tangent space $T_\gamma M$ 
for any curve f\/low $\gamma(t,x)$ in $M=G/SO(N)$ is def\/ined 
via 
\begin{gather}\label{ewrelation}
\covder{x} \coframe{} = 
[\coframe{},\gamma_{x}\hook \conx{}{}] 
\qquad\eqtext{ and }\qquad
\covder{t} \coframe{} = 
[\coframe{},\gamma_{t} \hook\conx{}{}] . 
\end{gather}
These derivatives $\covder{x}$, $\covder{t}$ obey the Cartan
structure equations \eqrefs{torseq}{curveq}, 
namely they have zero torsion
\begin{gather}\label{cartantors}
0 = (\covder{x} \gamma_{t} - \covder{t} \gamma_{x})\hook \coframe{} =
\D{x}\tframe{} - \D{t}\xframe{} 
+ [\xconx{}{},\tframe{}] -[\tconx{}{},\xframe{}]
\end{gather}
and carry $G$-invariant curvature
\begin{gather}
\label{cartancurv}
\curv{}{}(\gamma_{x},\gamma_{t}) \coframe{}
=
[\covder{x},\covder{t}] \coframe{}
= \D{x} \tconx{}{} - \D{t} \xconx{}{} + [\xconx{}{},\tconx{}{}] 
=-[\xframe{},\tframe{}],
\end{gather}
where
\begin{gather*} 
\xframe{}:= \gamma_{x} \hook \coframe{} ,\qquad
\tframe{}:= \gamma_{t} \hook \coframe{} ,\qquad
\xconx{}{}:= \gamma_{x} \hook \conx{}{} ,\qquad
\tconx{}{}:= \gamma_{t} \hook \conx{}{} .
\end{gather*}
The $G$-invariant covariant derivative and curvature on $T_\gamma M$
are thus seen to coincide with the Riemannian ones determined
from the metric $g$.
More generally, 
in this manner \cite{Sharpe} 
the relations~\eqrefs{kleinconx}{kleinframe}
canonically solder a Klein geometry onto a Riemannian symmetric-space geometry.

Geometrically, 
$\xframe{}$ and $\xconx{}{}$ represent the tangential part of 
the coframe and the connection $1$-form along $\gamma$.
For a non-stretching curve $\gamma$, 
where $x$ is the arclength, 
note $\xframe{}$ has unit norm in the inner product, 
$\langle \xframe{},\xframe{}\rangle_{\vs{p}}=1$.
This implies $\vs{p}$ has a decomposition into 
tangential and normal subspaces 
$\vs{p}_{\parallel}$ and $\vs{p}_{\perp}$ with respect
to $\xframe{}$ such that $\langle \xframe{},\vs{p}_{\perp}\rangle_{\vs{p}}=0$,
with $\vs{p}=\vs{p}_{\perp} \oplus \vs{p}_{\parallel}$ and
$\vs{p}_{\parallel} \simeq \Rnum{}$.

\begin{remark}
A main insight now, 
generalizing the results in \cite{SandersWang2,Anco}, 
is that the Cartan structure equations 
on the surface swept out by the curve f\/low $\gamma(t,x)$ in $M=G/SO(N)$
will geometrically encode $O(N-1)$-invariant
bi-Hamiltonian operators if the gauge (rotation) freedom of the
group action $SO(N)$ on $\coframe{}$ and $\conx{}{}$ is used to f\/ix
them to be a parallel coframe and its associated connection $1$-form
related by the Riemannian covariant derivative. 
The groups $G=SO(N+1)$ and $G=SU(N)$ will produce a dif\/ferent encoding 
except when $N=2$, since in that case 
$T_x M \simeq \vs{so}(3)/\vs{so}(2) \simeq \vs{su}(2)/\vs{so}(2)$ 
is the same tangent space for $M=SO(3)/SO(2)$ and $M=SU(2)/SO(2)$ 
due to the Lie-algebra isomorphism $\vs{so}(3) \simeq \vs{su}(2)$.
This will be seen to account for the existence of the two dif\/ferent
vector generalizations of the scalar mKdV hierarchy.
\end{remark}

The algebraic characterization of a parallel frame for curves 
in Riemannian geometry extends naturally to the setting of Klein geometry,
via the property that  
$\xframe{}$ is preserved by a $SO(N-1)$ rotation subgroup of 
the local frame structure group $SO(N)$
acting on $\vs{p} \subset \vs{g}$, 
while $\xconx{}{}$ belongs to the orthogonal complement of 
the $SO(N-1)$ rotation Lie subalgebra $\vs{so}(N-1)$ 
contained in the Lie algebra $\vs{so}(N)$ of $SO(N)$.
Their geometrical meaning, however,  
generalizes the Riemannian properties 
\eqrefs{parallelconxtang}{parallelconxperp},
as follows. 
Let $\frame{a}{}$ be a frame whose dual coframe 
is identif\/ied 
with the $\vs{p}$-valued coframe $\coframe{}$ 
in a f\/ixed orthonormal basis for $\vs{p}\subset \vs{g}$.
Decompose the coframe into parallel/perpendicular parts 
with respect to $\xframe{}$ in an algebraic sense 
as def\/ined by 
the kernel/cokernel of Lie algebra multiplication 
$[\xframe{},\cdot\ ]_\vs{g}=\ad(\xframe{}{})$.
Thus we have
$\coframe{}=(\coframe{C},\coframe{C^\perp})$
where the $\vs{p}$-valued covectors $\coframe{C}$, $\coframe{C^\perp}$ satisfy
$[\xframe{},\coframe{C}]_\vs{g}=0$,
and 
$\coframe{C^\perp}$ is orthogonal to $\coframe{C}$, 
so 
$[\xframe{},\coframe{C^\perp}]_\vs{g} \neq0$. 
Note there is a corresponding algebraic decomposition of
the tangent space $T_x M \simeq \vs{p}=\vs{g}/\vs{so}(N)$ given by 
$\vs{p}=\vs{p}_C \oplus \vs{p}_{C^\perp}$, with 
$\vs{p}_\parallel \subseteq \vs{p}_C$
and 
$\vs{p}_{C^\perp} \subseteq \vs{p}_\perp$,
where $[\vs{p}_\parallel,\vs{p}_C]=0$
and $\langle \vs{p}_{C^\perp},\vs{p}_C\rangle _\vs{p}=0$,
so $[\vs{p}_\parallel,\vs{p}_{C^\perp}]\neq 0$
(namely, $\vs{p}_C$ is the centralizer of $\xframe{}$ 
in $\vs{p} \subset \vs{g}$).
This decomposition is preserved by $\ad(\xconx{}{})$ 
which acts as an infinitesimal rotation,
$\ad(\xconx{}{})\vs{p}_C \subseteq \vs{p}_{C^\perp}$, 
$\ad(\xconx{}{})\vs{p}_{C^\perp} \subseteq \vs{p}_C$. 
Hence, from equation \eqref{ewrelation},
the derivative $\covder{x}$ of a covector perpendicular
(respectively parallel) to $\xframe{}$ 
lies parallel (respectively perpendicular) to $\xframe{}$,
namely 
$\covder{x}\coframe{C}$ belongs to $\vs{p}_{C^\perp}$,
$\covder{x}\coframe{C^\perp}$ belongs to $\vs{p}_C$. 
In the Riemannian setting,
these properties correspond to 
$\gcovder{x}(\frame{}{a})_C = \v{a}{\ b}(\frame{}{b})_{C^\perp}$, 
$\gcovder{x}(\frame{}{a})_{C^\perp} = -\v{\ a}{b}(\frame{}{b})_C$
for some functions $\v{ab}{}=-\v{ba}{}$.
Such a frame will be called {\it $SO(N)$-parallel}
and def\/ines a strict generalization of a~Riemannian parallel frame
whenever $\vs{p}_C$ is larger than $\vs{p}_\parallel$.

Existence of a $SO(N)$-parallel frame for curve flows in
Klein geometries $G/SO(N)$ is guaranteed by the $SO(N)$ gauge freedom
on $\frame{}{}$ and $\conx{}{}$ inherited from
the local section of $G$ used to pull back the Maurer-Cartan form
to $G/SO(N)$.

\section[Bi-Hamiltonian operators and vector soliton equations for
$SO(N+1)/SO(N) \simeq S^N$]{Bi-Hamiltonian operators and vector soliton equations\\ for
$\boldsymbol{SO(N+1)/SO(N) \simeq S^N}$}

Recall
$\vs{so}(k)$ is a real vector space 
isomorphic to the Lie algebra of $k\times k$ skew-symmetric matrices. 
So the tangent space $T_x M= \vs{so}(N+1)/\vs{so}(N)$ 
of the Riemannian manifold $M=SO(N+1)/SO(N)$
is isomorphic to $\vs{p} \simeq \Rnum{N}$,
as described by the following canonical decomposition 
\begin{gather*}
\begin{pmatrix}
0 & p \\ -\trans{p} & \bdsymb{0}
\end{pmatrix}
\in \vs{p} \subset \vs{so}(N+1)=\vs{g}  ,\qquad
\bdsymb{0} \in \vs{so}(N) =\vs{h} ,\qquad
p \in \Rnum{N}
\end{gather*}
parameterized by the $N$ component vector $p$. 
The Cartan--Killing inner product on $\vs{g}$
is given by the trace of the product of 
an $\vs{so}(N+1)$ matrix and a transpose $\vs{so}(N+1)$ matrix, 
multiplied by a normalization factor
$\frac{1}{2}$. 
The norm-squared on the quotient space $\vs{p}$ 
thereby reduces to the ordinary dot product of vectors $p$:
\begin{gather*} 
\left< 
\begin{pmatrix}0 & p\\ -\trans{p} &\bdsymb{0} \end{pmatrix}, 
\begin{pmatrix}0 & p\\ -\trans{p} &\bdsymb{0} \end{pmatrix} 
\right> 
= \frac{1}{2} \tr\left( 
\trans{\begin{pmatrix}0 & p\\ -\trans{p} & \bdsymb{0} \end{pmatrix}} 
\begin{pmatrix}0 & p\\ -\trans{p} & \bdsymb{0} \end{pmatrix} 
\right)
= p \cdot p .
\end{gather*}
Note the Cartan--Killing inner product provides a canonical
identif\/ication between $\vs{p} \simeq \Rnum{N}$ and its dual
$\vs{p}^{*} \simeq \Rnum{N}$.

Let $\gamma(t,x)$ be a f\/low of a non-stretching curve in 
$M =SO(N+1)/SO(N) \simeq S^N$.
We introduce a $SO(N)$-parallel coframe 
$\coframe{} \in T^*_\gamma M\otimes\vs{p}$ 
and its associated connection $1$-form
$\conx{}{} \in T^*_\gamma M\otimes\vs{so}(N)$ 
along $\gamma$
by putting\footnote{Note $\conx{}{}$ is related to $\coframe{}$ by
the Riemannian covariant derivative \eqref{ewrelation}
on the surface swept out by the curve f\/low~$\gamma(t,x)$.}
\begin{gather} 
\xframe{} 
= \gamma_{x} \hook \coframe{} 
= \begin{pmatrix}
0 & (1,\vec{0}) \\ -\trans{(1,\vec{0})} & \bdsymb{0} 
\end{pmatrix} 
\in \vs{p} ,\qquad
(1,\vec{0}) \in \Rnum{N} ,\qquad
\vec{0} \in \Rnum{N-1}
\end{gather} 
and
\begin{gather*} 
\xconx{}{} 
= \gamma_{x} \hook \conx{}{} 
= \begin{pmatrix}
0 & (0,\vec{0}) \\ 
\trans{(0,\vec{0})} & \bdsymb{\xconx{}{}}
\end{pmatrix} 
\in \vs{so}(N+1),
\end{gather*}
where
\begin{gather*}
\bdsymb{\xconx{}{}} = 
\begin{pmatrix}
0 & \vvec \\ -\trans{\vvec} & \bdsymb{0} 
\end{pmatrix} 
\in \vs{so}(N) ,\qquad
\vvec \in \Rnum{N-1} ,\qquad
\bdsymb{0} \in \vs{so}(N-1) .
\end{gather*}
Note the form of $\xframe{}$ indicates the coframe $\coframe{}$ is
adapted to $\gamma$, with $(1,\vec{0})$ representing a choice of a
constant unit-norm vector in $\vs{p} \simeq \Rnum{N}$, so
$\langle \xframe{},\xframe{}\rangle_{\vs{p}} 
= (1,\vec{0}) \cdot (1,\vec{0}) =1$.
All such choices are equivalent through the $SO(N)$ rotation gauge
freedom on the coframe and connection $1$-form.  
Consequently, there is a decomposition of $SO(N+1)/SO(N)$ matrices
\begin{gather*} 
\begin{pmatrix}
0 & p \\ -\trans{p} & \bdsymb{0} 
\end{pmatrix} 
=
\begin{pmatrix}
0 & (p_{\parallel},\vec{0}) \\
-\trans{(p_{\parallel},\vec{0})} & \bdsymb{0} 
\end{pmatrix} 
+
\begin{pmatrix}0 & (0,\vec{p}_{\perp}) \\
-\trans{(0,\vec{p}_{\perp})} & \bdsymb{0} \end{pmatrix} 
\in \vs{p}
\end{gather*}
into tangential and normal parts relative to $\xframe{}$ via a
corresponding decomposition of vectors given by
\begin{gather*} 
p = (p_{\parallel},\vec{p}_{\perp}) \in \Rnum{N} 
\end{gather*}
relative to $(1,\vec{0})$.
Thus $p_{\parallel}$ is identif\/ied with $\vs{p}_\parallel=\vs{p}_C$,
and $\vec{p}_{\perp}$ with $\vs{p}_\perp=\vs{p}_{C^\perp}$.

In the f\/low direction we put
\begin{gather*} 
\tframe{} = \gamma_{t} \hook \coframe{} 
= \begin{pmatrix}
0 & (\hpar,\vec{h}_{\perp}) \\
-\trans{(\hpar,\vec{h}_{\perp})} & \bdsymb{0} 
\end{pmatrix} 
\in \vs{p} ,\qquad
(\hpar,\vec{h}_{\perp}) \in \Rnum{N} ,\qquad
\vec{h}_{\perp} \in \Rnum{N-1}
\end{gather*}
and
\begin{gather}\label{soflowconx}
\tconx{}{} = \gamma_{t} \hook \conx{}{} 
= 
\begin{pmatrix}
0 & (0,\vec{0}) \\ \trans{(0,\vec{0})} & \bdsymb{\tconx{}{}} 
\end{pmatrix} 
\in \vs{so}(N+1), 
\end{gather}
where
\begin{gather}\label{soflowconxh}
\bdsymb{\tconx{}{}}  = 
\begin{pmatrix}
0 & \wvec \\ -\trans{\wvec} & \bdsymb{\Theta} 
\end{pmatrix} 
\in \vs{so}(N) ,\qquad
\wvec \in \Rnum{N-1} ,\qquad
\bdsymb{\Theta} \in \vs{so}(N-1) .
\end{gather}
The components $\hpar$, $\vec{h}_{\perp}$
correspond to decomposing 
$\tframe{} =
g(\gamma_{t},\gamma_{x})\xframe{}+(\gamma_{t})_{\perp} \hook \coframe{\perp}$ 
into tangential and normal parts relative to $\xframe{}$.  
We now have
\begin{gather*}
[\xframe{},\tframe{}] = 
-\begin{pmatrix}
0 & 0 \\ 0 & \bdsymb{h_{\perp}} 
\end{pmatrix} 
\in \vs{so}(N+1) ,\qquad
\bdsymb{h_{\perp}} =
\begin{pmatrix}
0 & \hvec_{\perp} \\ -\trans{\hvec_{\perp}} & \bdsymb{0} \end{pmatrix} 
\in \vs{so}(N  ) ,
\\
[\tconx{}{},\tframe{}] = -
\begin{pmatrix}
0 & (0,\wvec) \\ -\trans{(0,\wvec)} & 0 
\end{pmatrix} 
\in \vs{p}_{\perp} ,
\\ 
[\xconx{}{},\tframe{}] = 
-\begin{pmatrix}
0 & (-\vvec \cdot \hvec_{\perp}, \hpar \vvec) \\ 
-\trans{(-\vvec \cdot\hvec_{\perp}, \hpar \vvec)} & \bdsymb{0} 
\end{pmatrix}
\in \vs{p} .
\end{gather*}
Hence the curvature equation \eqref{cartancurv} reduces to
\begin{gather}
\D{t} \vvec - \D{x} \wvec - \vvec \hook \bdsymb{\Theta} =
-\hvec_{\perp} ,
\label{soveq}\\
-\D{x} \bdsymb{\Theta} + \vvec \otimes \wvec -\wvec \otimes \vvec =0 ,
\label{sothetaeq}
\end{gather}
while the torsion equation \eqref{cartantors} yields
\begin{gather}
\label{sohpareq}
0 = \D{x} \hpar + \vvec \cdot \hvec_{\perp} ,
\\
\label{soweq}
0 = \wvec - \hpar \vvec + \D{x} \hvec_{\perp} .
\end{gather} 
Here $\otimes$ denotes the outer product of pairs of vectors
($1 \times N$ row matrices), 
producing $N \times N$ matrices 
$\vec{A} \otimes \vec{B} = \trans{\vec{A}} \vec{B}$, 
and $\hook$ denotes
multiplication of $N \times N$ matrices 
on vectors ($1 \times N$ row matrices), 
$\vec{A} \hook (\vec{B} \otimes \vec{C}) := (\vec{A} \cdot \vec{B}) \vec{C}$ 
which is the transpose of the standard matrix product on column vectors, 
$(\vec{B} \otimes \vec{C}) \vec{A} = (\vec{C} \cdot \vec{A}) \vec{B}$.  
To proceed we use equations \eqrefs{sothetaeq}{sohpareq} to eliminate
\begin{gather}\label{sotheta}
\bdsymb{\Theta} 
= -\Dinv{x}(\wvec \otimes \vvec -\vvec \otimes \wvec) ,\qquad
\hpar 
= -\Dinv{x} (\vvec \cdot \hvec_{\perp}) 
\end{gather}
in terms of the variables $\vvec$, $\hvec_{\perp}$, $\wvec$.
Then equation \eqref{soveq} gives a f\/low on $\vvec$,
\begin{gather*} 
\vvec_{t} 
= \D{x} \wvec - \vvec \hook \Dinv{x}(
\wvec \otimes \vvec - \vvec \otimes \wvec) - \chi \hvec_{\perp}
\end{gather*}
with
\begin{gather*} 
\wvec = -\Dinv{x} (\vvec \cdot \hvec_{\perp}) \vvec - \D{x}\hvec_{\perp} 
\end{gather*}
obtained from equation \eqref{soweq}.  
Here $\chi=1$ represents the Riemannian
scalar curvature of $SO(N+1)/SO(N) \simeq S^{N}$ (see \cite{Anco}).
In these equations we read of\/f the operators
\begin{gather*} 
\Hop = \D{x} + \vvec \hook \Dinv{x}(\vvec \wedge\ ) ,\qquad
\Jop = \D{x} + \Dinv{x}(\vvec \cdot\ ) \vvec ,
\end{gather*}
where 
$\vec{A} \wedge \vec{B} = \vec{A} \otimes \vec{B} - \vec{B} \otimes \vec{A}$.
The results in \cite{SandersWang2} prove the following properties of 
$\Hop$, $\Jop$.

\begin{theorem}\label{thm1} 
$\Hop$, $\Jop$ are compatible $O(N-1)$-invariant
Hamiltonian cosymplectic and symplectic operators with respect to
the Hamiltonian variable $\vvec$.  
Consequently, the flow equation
takes the Hamiltonian form
\begin{gather*} 
\vvec_{t} = \Hop(\wvec) - \chi \hvec_{\perp} =
\Rop(\hvec_{\perp}) - \chi \hvec_{\perp} ,\qquad
\wvec = \Jop(\hvec_{\perp})
\end{gather*}
where $\Rop = \Hop \circ \Jop$ is a hereditary recursion operator.
\end{theorem}

On the $x$-jet space of $\vvec(t,x)$, the variables $\hvec_{\perp}$
and $\wvec$ have the respective meaning of a Hamiltonian vector f\/ield 
$\hvec_{\perp}\hook \partial/\partial \vvec$ 
and covector f\/ield $\wvec\hook d\vvec$
(see the Appendix of \cite{Anco}).
Thus the recursion operator\footnote{This $O(N-1)$-invariant form of the recursion operator 
appeared already in \cite{Anco2}.}
\begin{gather}
\Rop = 
\D{x}(\D{x} + \Dinv{x}(\vvec \cdot\ ) \vvec) 
+ \vvec \hook \Dinv{x}(\vvec \wedge \D{x}\ ) 
\nonumber\\ 
\phantom{\Rop}{} = 
\D{x}^2 +|\vvec|^2 +\Dinv{x}(\vvec\cdot\ )\vvec_x 
-\vvec \hook \Dinv{x}(\vvec_x \wedge\ )\label{soRop}
\end{gather}
generates a hierarchy of commuting Hamiltonian vector f\/ields
$\hvec^{(k)}_{\perp}$ starting from $\hvec^{(0)}_{\perp}=\vvec_{x}$
given by the inf\/initesimal generator of $x$-translations in terms of
arclength $x$ along the curve.

The adjoint operator $\Rop^*$ generates a related hierarchy of
involutive Hamiltonian covector f\/ields 
$\wvec^{(k)} = \delta H^{(k)}/\delta \vvec$ 
in terms of Hamiltonians 
$H =H^{(k)}(\vvec,\vvec_{x},\vvec_{2x},\ldots)$ starting from
$\wvec^{(0)}=\vvec$, $H^{(0)}=\frac{1}{2} |\vvec|^{2}$.  
These hierarchies are
related by $\hvec^{(k)}_{\perp} = \Hop(\wvec^{(k)})$,
$\wvec^{(k+1)}=\Jop(\hvec^{(k)}_{\perp})$, $k=0,1,2,\ldots$.
Both hierarchies share the mKdV scaling symmetry
$x\rightarrow\lambda x$, $\vvec\rightarrow\lambda^{-1}\vvec$, 
under which we see
$\hvec^{(k)}_{\perp}$ and $H^{(k)}$ have scaling weight $2+2k$,
while $\wvec^{(k)}$ has scaling weight $1+2k$.

\begin{corollary}\label{cor1}
Associated to the recursion operator $\Rop$ there is a
corresponding hierarchy of commuting bi-Hamiltonian flows on $\vvec$
given by $O(N-1)$-invariant vector evolution equations
\begin{gather}\label{floweq}
\vvec_{t} = \hvec^{(k+1)}_{\perp} - \chi \hvec^{(k)}_{\perp} 
=\Hop(\delta H^{(k,\chi)}/\delta \vvec) 
= \Jop^{-1}(\delta H^{(k+1,\chi)}/\delta \vvec) ,\qquad
k = 0,1,2,\ldots 
\end{gather}
with Hamiltonians $H^{(k+1,\chi)} = H^{(k+1)} -\chi H^{(k)}$,
where $\Hop,\Jop^{-1}$ are compatible Hamiltonian opera\-tors.
An alternative (explicit) Hamiltonian operator for these flows is
$\Eop:= \Hop\circ\Jop\circ\Hop=\Rop\circ\Hop$. 
\end{corollary}

\begin{remark}\label{rem1}
Using the terminology of \cite{AncoWolf},
$\hvec^{(k)}_{\perp}$ will be said to produce a $+(k+1)$ f\/low on $\vvec$.
This dif\/fers from the terminology in \cite{Anco}
which refers to equation \eqref{floweq} 
as the $+k$ f\/low.
\end{remark}

The $+1$ f\/low given by $\hvec_{\perp} = \vvec_{x}$ yields
\begin{gather}\label{somkdveq}
\vvec_{t} 
= \vvec_{3x} + \frac{3}{2} |\vvec|^{2} \vvec_{x} - \chi \vvec_{x}
\end{gather}
which is a vector mKdV equation up to a convective term that can be
absorbed by a Galilean transformation $x \rightarrow x - \chi t$, $t
\rightarrow t$.
The $+(k+1)$ f\/low gives a vector mKdV equation of higher order $3 + 2k$
on $\vvec$.

There is also a $0$ f\/low $\vvec_{t} = \vvec_{x}$ arising from
$\hvec_{\perp}=0$, $\hpar=1$, which falls outside the hierarchy
generated by $\Rop$.

All these f\/lows correspond to geometrical motions of the curve $\gamma(t,x)$,
given by
\begin{gather*}
\gamma_{t}=
f(\gamma_{x},\covder{x}\gamma_{x},\covder{x}^{2}\gamma_{x},\ldots)
\end{gather*}
subject to the non-stretching condition 
\begin{gather*}
|\gamma_{x}|_{g}=1 .
\end{gather*}
The equation of motion is obtained 
from the identif\/ications
$\gamma_{t} \leftrightarrow \tframe{}$, 
$\covder{x}\gamma_{x} \leftrightarrow \covD{x}\xframe{} 
= [\xconx{}{},\xframe{}]$, 
and so on, using 
$\covder{x} \leftrightarrow \D{x} + [\xconx{}{},\cdot] = \covD{x}$. 
These identif\/ications correspond to $T_x M \leftrightarrow \vs{p}$
as def\/ined via the parallel coframe. 
Note we have
\begin{gather*} 
[\xconx{}{},\xframe{}] = 
-\begin{pmatrix}
0 & (0,\vvec)\\
-\trans{(0,\vvec)} & \bdsymb{0} 
\end{pmatrix} ,
\\
[\xconx{}{},[\xconx{}{},\xframe{}]] = 
-\begin{pmatrix}
0 & (|\vvec|^{2},\vec{0})\\ -\trans{(|\vvec|^{2},\vec{0})} & \bdsymb{0} 
\end{pmatrix} 
= -|\vvec|^{2} \xframe{} ,
\end{gather*}
and so on.  
In particular, for the $+1$ f\/low,
\begin{gather*} 
\hvec_{\perp} = \vvec_x ,\qquad
\hpar = -\Dinv{x}(\vvec \cdot \vvec_{x}) 
=-\frac{1}{2}|\vvec|^{2} ,
\end{gather*}
thus
\begin{gather*}
\tframe{} 
=
\begin{pmatrix}
0 & (\hpar,\vec{h}_{\perp}) \\
-\trans{(\hpar,\vec{h}_{\perp})} & \bdsymb{0} 
\end{pmatrix} 
= -\frac{1}{2}|\vvec|^2 
\begin{pmatrix}
0 & (1,\vec{0})\\ -\trans{(1,\vec{0})} & \bdsymb{0} 
\end{pmatrix} 
+ 
\begin{pmatrix}
0 & (0,\vvec_x)\\ -\trans{(0,\vvec_x)} & \bdsymb{0} \end{pmatrix} 
\\ \nonumber 
\phantom{\tframe{}}{}=
-\D{x}[\xconx{}{},\xframe{}]+\frac{1}{2}[\xconx{}{},[\xconx{}{},\xframe{}]]
= 
-\covD{x}[\xconx{}{},\xframe{}]-\frac{3}{2}|\vvec|^2\xframe{} .
\nonumber
\end{gather*}
We identify the f\/irst term as
$-\covder{x}(\covder{x}\gamma_x)=-\covder{x}^2\gamma_x$.
For the second term we observe
$|\vvec|^2= \langle [\xconx{}{},\xframe{}],\xconx{}{},\xframe{}]\rangle_{\vs{p}}
\leftrightarrow
g(\covder{x}\gamma_x,\covder{x}\gamma_x)=|\covder{x}\gamma_x|_{g}^{2}$
since the Cartan--Killing inner product corresponds to the Riemannian
metric $g$.
Hence we have $\tframe{} \leftrightarrow -(\covder{x}^2\gamma_x +
\frac{3}{2}|\covder{x}\gamma_x|_{g}^{2} \gamma_x)$.
This describes a geometric nonlinear PDE for $\gamma(t,x)$,
\begin{gather}\label{mkdvmap}
-\gamma_t = 
\covder{x}^2\gamma_x +\frac{3}{2}|\covder{x}\gamma_x|_{g}^{2} \gamma_x 
\end{gather}
which is referred to as the {\it non-stretching mKdV map equation} on the
symmetric space $M = SO(N+1)/SO(N) \simeq S^N$.
A dif\/ferent derivation using just the Riemannian geometry of $S^N$
was given in \cite{Anco}.
Since in the tangent space $T_x S^N\simeq \vs{so}(N+1)/\vs{so}(N)$
the kernel of $[\xframe{},\cdot\ ]$ is spanned by $\xframe{}$,
a parallel frame in the setting of the Klein geometry of $SO(N+1)/SO(N)$ 
is precisely the same as in the Riemannian geometry of $S^N$.

The convective term $|\covder{x}\gamma_x|_{g}^{2} \gamma_x$ can be
written in an alternative form using the Klein geometry of 
$SO(N+1)/SO(N) \simeq S^N$. 
Let $\ad(\cdot)$ denote the standard adjoint representation acting in the
Lie algebra $\vs{g}=\vs{p} \oplus \vs{so}(N)$.
We f\/irst observe
\begin{gather*} 
\ad([\xconx{}{},\xframe{}])\xframe{} = 
\begin{pmatrix}
0 & (0,\vec{0})\\ \trans{(0,\vec{0})} & \bdsymb{v} 
\end{pmatrix} 
\in \vs{so}(N+1) ,
\end{gather*}
where
\begin{gather*} 
\bdsymb{v} = 
-\begin{pmatrix}
0 & \vvec\\ -\trans{\vvec} & \bdsymb{0} 
\end{pmatrix} 
\in \vs{so}(N) .
\end{gather*}
Applying $\ad([\xconx{}{},\xframe{}])$ again, we obtain
\begin{gather*} 
\ad([\xconx{}{},\xframe{}])^2\xframe{}
= -|\vvec|^2 
\begin{pmatrix}
0 & (1,\vec{0})\\ -\trans{(1,\vec{0})} &\bdsymb{0} 
\end{pmatrix} 
= -|\vvec|^2 \xframe{} .
\end{gather*}
Hence, 
$|\vvec|^2 \xframe{} \leftrightarrow
-\chi^{-1}\ad(\covder{x}\gamma_x)^2 \gamma_x 
= |\covder{x} \gamma_x|_{g}^{2} \gamma_x$, 
and thus the mKdV map equation is equivalent to
\begin{gather}\label{symmspmkdvmap}
-\gamma_t = 
\covder{x}^2 \gamma_x 
- \frac{3}{2}\chi^{-1}\ad(\covder{x} \gamma_x)^2 \gamma_x .
\end{gather}
Note here that $\ad(\covder{x} \gamma_x) = [\covder{x}\gamma_x,\cdot\ ]$
maps $\vs{p} \simeq T_x M$ into $\vs{so}(N)$ 
and maps $\vs{so}(N)$ into $\vs{p} \simeq T_x M$, 
so $\ad(\covder{x}\gamma_x)^2$ is well-def\/ined on the
tangent space $T_x M \simeq \vs{p}$ of $M=SO(N+1)/SO(N)$.

Higher f\/lows on $\vvec$ yield higher-order geometric PDEs. 
The $0$ f\/low on $\vvec$ directly corresponds to 
\begin{gather}\label{convmap}
\gamma_t=\gamma_x
\end{gather}
which is just a convective (linear traveling wave) map equation.

There is a $-1$ f\/low contained in the hierarchy,
with the property that $\hvec_{\perp}$ is annihilated by 
the symplectic operator $\Jop$
and hence gets mapped into $\Rop(\hvec_{\perp})=0$ 
under the recursion operator. 
Geometrically this f\/low means simply 
$\Jop(\hvec_{\perp})=\wvec=0$
which implies $\tconx{}{}=0$ 
from equations~\eqref{soflowconx}, \eqref{soflowconxh}, \eqref{sotheta},
and hence 
$0=[\tconx{}{},\xframe{}]=\covD{t}\xframe{}$ 
where 
$\covD{t} = \D{t}+ [\tconx{}{},\cdot]$.
The correspondence $\covder{t} \leftrightarrow \covD{t}$, $\gamma_x
\leftrightarrow \xframe{}$ immediately leads to the equation of motion
\begin{gather}\label{wavemap}
0 = \covder{t}\gamma_x 
\end{gather}
for the curve $\gamma(t,x)$.
This nonlinear geometric PDE is precisely a wave map equation on the
symmetric space $SO(N+1)/SO(N) \simeq S^N$.
The resulting f\/low equation on $\vvec$ is 
\begin{gather}\label{sovsgflow}
\vvec_t = -\chi \hvec_{\perp} ,\qquad \chi=1,
\end{gather}
where 
\begin{gather*}
0=\wvec = -\D{x}\hvec_{\perp} + \hpar \vvec ,\qquad
\D{x}\hpar = -\hvec_{\perp}\cdot\vvec .
\end{gather*}
Note this f\/low equation possesses the conservation law 
$0=\D{x}( \hpar{}^2+|\hvec_{\perp}|^2 )$
with
\begin{gather*}
\hpar{}^2+|\hvec_{\perp}|^2 = \langle\tframe{},\tframe{}\rangle_\vs{p} = |\gamma_t|_g^2
\end{gather*}
corresponding to 
\begin{gather}\label{wavemapconslaw}
0=\covder{x}|\gamma_t|_g^2 .
\end{gather}
Thus a conformal scaling of $t$ can be used to put 
$|\gamma_t|_g=1$, and so
\begin{gather*}
1 = \hpar{}^2+|\hvec_{\perp}|^2 .
\end{gather*}
Substitution of $\hpar=\sqrt{1-|\hvec_{\perp}|^2}$
along with $\hvec_{\perp} = -\chi^{-1}\vvec_t$
into the equation $\D{x}\hvec_{\perp} =\hpar \vvec$
consequently reduces the wave map equation to 
a hyperbolic vector equation
\begin{gather}\label{sosghyperboliceq}
\vvec_{tx} = 
-\sqrt{\chi^2-|\vvec_{t}|^2} \vvec ,\qquad \chi=1. 
\end{gather}
Equivalently, $\vvec$ satisfies a nonlocal evolution equation
\begin{gather*}
\vvec_{t} = 
-\Dinv{x}\Big(\sqrt{1-|\vvec_{t}|^2} \vvec\Big) 
\end{gather*}
describing the $-1$ f\/low. 
It also follows from 
$\vvec = \hpar^{-1}\D{x}\hvec_\perp$
combined with the flow equation \eqref{sovsgflow}
that $\hvec_{\perp}$ obeys the vector $SG$ equation
\begin{gather}\label{soSGeq}
\Big(\sqrt{(1-|\hvec_{\perp}|^2)^{-1}} \hvec_{\perp x}\Big){}_t 
= -\hvec_{\perp} 
\end{gather}
which has been derived previously in \cite{Bakas,Pohlmeyer,Wang}
from a dif\/ferent point of view. 
These equations~\eqrefs{sosghyperboliceq}{soSGeq}
possess the mKdV scaling symmetry
$x\rightarrow\lambda x$, $\vvec\rightarrow\lambda^{-1}\vvec$, 
where $\hvec_{\perp}$ has scaling weight $0$.

The hyperbolic vector equation \eqref{sosghyperboliceq}
admits the vector mKdV equation \eqref{somkdveq}
as a higher symmetry, 
which is shown by the symmetry-integrability classif\/ication 
results in \cite{AncoWolf}.
As a~consequence of Corollary~\ref{cor1}, we see that 
the recursion operator $\Rop=\Hop\circ\Jop$ 
generates a hierarchy of vector mKdV symmetries
\begin{gather} 
\vvec_{t}^{(0)} 
= \vvec_x ,
\label{so0flow}\\  
\vvec_{t}^{(1)} 
 = \Rop(\vvec_x) 
= \vvec_{3x} + \frac{3}{2}|\vvec|^2 \vvec_x ,
\label{so1flow}\\
\vvec_{t}^{(2)} 
= \Rop^{2}(\vvec_x) 
= \vvec_{5x} +\frac{5}{2}(|\vvec|^2\vvec_{2x})_x 
+ \frac{5}{2}\left((|\vvec|^2)_{xx} +|\vvec_x|^2 + \frac{3}{4}|\vvec|^4\right)\vvec_x 
- \frac{1}{2}|\vvec_x|^2 \vvec ,
\label{so2flow}
\end{gather}
and so on, all of which commute with the $-1$ f\/low
\begin{gather}\label{so-1flow}
\vvec_{t}^{(-1)}
=\hvec_{\perp} 
\end{gather}
associated to the vector SG equation \eqref{soSGeq}.  
Moreover the adjoint operator $\Rop^* =\Jop\circ\Hop$ 
generates a hierarchy of mKdV Hamiltonians
\begin{gather*} 
H^{(0)} = \frac{1}{2}|\vvec|^2 ,
\\  
H^{(1)} = 
-\frac{1}{2}|\vvec_x|^2+\frac{1}{8}|\vvec|^4 ,
\\  
H^{(2)} =
\frac{1}{2}|\vvec_{2x}|^2-\frac{3}{4}|\vvec|^2|\vvec_x|^2
-\frac{1}{2}(\vvec\cdot \vvec_x)^2 + \frac{1}{16}|\vvec|^6 ,
\end{gather*}
and so on, all of which are conserved densities for the $-1$ f\/low.
It f\/ollows that 
the hyperbolic vector equations \eqrefs{sosghyperboliceq}{soSGeq}
admit these respective hierarchies 
of vector mKdV symmetries and conserved densities. 

Viewed as f\/lows,
the entire hierarchy of vector PDEs 
\eqref{so-1flow}, \eqsref{so0flow}{so2flow}, etc. 
possesses the mKdV scaling symmetry 
$x\rightarrow\lambda x$, $\vvec\rightarrow\lambda^{-1}\vvec$, 
with $t\rightarrow\lambda^{1+2k} t$
for $k=-1,0,1,2$, etc.
Moreover for $k\geq 0$,
all these expressions will be local polynomials in the variables
$\vvec,\vvec_x,\vvec_{xx},\ldots$
as established by general results in \cite{Wang-thesis,Sergyeyev}
concerning nonlocal recursion operators. 

\begin{theorem}\label{thm2}
In the symmetric space $SO(N+1)/SO(N)$
there is a hierarchy of bi-Hamiltonian flows of curves $\gamma(t,x)$
described by geometric map equations. 
The $0$ flow is a convective (traveling wave) map \eqref{convmap},
while the $+1$ flow is a non-stretching mKdV map \eqref{mkdvmap}
and the $+2,\ldots$ flows are higher order analogs. 
The kernel of the recursion operator \eqref{soRop} in the hierarchy 
yields the $-1$ flow which is a non-stretching wave map \eqref{wavemap}.
\end{theorem}

\section[Bi-Hamiltonian operators and vector soliton equations
for $SU(N)/SO(N)$]{Bi-Hamiltonian operators and vector soliton equations\\
for $\boldsymbol{SU(N)/SO(N)}$}

Recall $\vs{su}(k)$ is a complex vector space 
isomorphic to the Lie algebra of $k\times k$ skew-hermitian matrices.
The real and imaginary parts of these matrices respectively belong to 
the real vector space $\vs{so}(k)$ of skew-symmetric matrices
and the real vector space $\vs{s}(k) \simeq \vs{su}(k)/\vs{so}(k)$
def\/ined by $k\times k$ symmetric trace-free matrices. 
Hence $\vs{g}= \vs{su}(N)$ has the decomposition 
$\vs{g} =\vs{h} +\i \vs{p}$ 
where $\vs{h}=\vs{so}(N)$
and $\vs{p} =\vs{s}(N)$. 
The Cartan--Killing inner product is given by 
the trace of the product of an $\vs{su}(N)$ matrix 
and a hermitian-transpose $\vs{su}(N)$ matrix, 
multiplied by $1/2$. 
Note any matrix in $\vs{s}(N)$ can be diagonalized under the action
of the group $SO(N)$.

Let $\gamma(t,x)$ be  a f\/low of a non-stretching curve in $M=SU(N)/SO(N)$
where we identify $T_x M \simeq \vs{p}$
(dropping a factor $\i$ 
for simplicity\footnote{Retaining the $\i$ in this identif\/ication will change
only the sign of the scalar curvature factor $\chi$ in the f\/low equation.}).
We consider a $SO(N)$-parallel coframe 
$\coframe{} \in T^*_\gamma M\otimes\vs{p}$ 
and its associated connection $1$-form
$\conx{}{} \in T^*_\gamma M\otimes\vs{so}(N)$ along 
$\gamma$
given by\footnote{As before, $\conx{}{}$ is related to $\coframe{}$ by
the Riemannian covariant derivative \eqref{ewrelation}
on the surface swept out by the curve f\/low $\gamma(t,x)$.} 
\begin{gather} 
\xframe{} = \gamma_x \hook \coframe{}= 
\kappa\left( \begin{pmatrix}
-1 & \vec{0}\\ \trans{\vec{0}} & \bdsymb{0} 
\end{pmatrix} 
+ \frac{1}{N} \idmatr \right)
=\frac{\kappa}{N} 
\begin{pmatrix}
1-N & \vec{0}\\ \trans{\vec{0}} & \bdsymb{1} 
\end{pmatrix} 
\in \vs{p}  ,\label{suframe}\\
\bdsymb{0},\i\bdsymb{1} \in \vs{u}(N-1) ,\qquad 
\vec{0} \in \Rnum{N-1}\nonumber
\end{gather}
up to a normalization factor $\kappa$ which we will f\/ix shortly,
and
\begin{gather}\label{suconx} 
\xconx{}{} = 
\begin{pmatrix}
0 & \vvec\\ -\trans{\vvec} & \bdsymb{0} 
\end{pmatrix} 
\in \vs{so}(N) ,\qquad
\vvec \in \Rnum{N-1} .
\end{gather}
Since the form of $\xframe{}$ is a constant matrix, it indicates that
the coframe is adapted to $\gamma$ provided~$\xframe{}$ has unit
norm in the Cartan--Killing inner product.
We have 
\begin{gather}\label{normalize}
\langle \xframe{},\xframe{}\rangle_{\vs{p}} 
= \frac{\kappa^2}{2} \tr
\begin{pmatrix}(N^{-1}-1)^2 & 0\\ 0 & N^{-1}\bdsymb{1} \end{pmatrix} 
= \kappa^2 (N-1)/(2N)=1
\end{gather} 
after putting $\kappa^2=2 N(N-1)^{-1}$.
As a consequence, all matrices in $\vs{p}=\vs{s}(N)$ will
have a canonical decomposition into tangential and normal parts relative to
$\xframe{}$,
\begin{gather*}
\begin{pmatrix}
(N^{-1}-1) p_{\parallel} & \vec{p}_{\perp}\\ \trans{\vec{p}_{\perp}} &
\bdsymb{p_{\perp}}-N^{-1} p_{\parallel}\bdsymb{1} 
\end{pmatrix} 
=\frac{1}{N}
\begin{pmatrix}
(1-N) p_{\parallel} & \vec{0}\\ \trans{\vec{0}} & p_{\parallel}\bdsymb{1} 
\end{pmatrix} 
+ 
\begin{pmatrix}
0 &\vec{p}_{\perp}\\ \trans{\vec{p}_{\perp}} &\bdsymb{p_{\perp}} 
\end{pmatrix} 
\end{gather*}
parameterized by the $(N-1)\times (N-1)$ matrix 
$\bdsymb{p_{\perp}} \in \vs{s}(N-1)$
and the $N$ component vector 
$(p_{\parallel},\vec{p}_{\perp}) \in \Rnum{N}$, 
corresponding to the decomposition 
$\vs{s}(N) =
\vs{s}(N)_{\parallel} \oplus \vs{s}(N)_{\perp}$ 
given by 
$\langle \vs{s}(N)_{\perp},\xframe{}\rangle_{\vs{p}}=0$ 
and
$\langle\vs{s}(N)_{\parallel},\xframe{}\rangle_{\vs{p}}=\kappa p_{\parallel}$ 
under the previous normalization of $\xframe{}$.
Here $(p_\parallel,\bdsymb{p_\perp})$ is identif\/ied with 
$\vs{p}_C \supset \vs{p}_\parallel$,
and $\vec{p}_\perp$ with $\vs{p}_{C^\perp}\subset \vs{p}_\perp$.
Note $\bdsymb{p_\perp}$ is empty only if $N=2$,
so consequently for $N>2$ 
the $SO(N)$-parallel frame \eqrefs{suframe}{suconx}
is a strict generalization of a~Riemannian parallel frame. 

In the f\/low direction we put
\begin{gather}
\tframe{} = \gamma_{t} \hook \coframe{} 
= \kappa\left( \hpar
\begin{pmatrix}
N^{-1}-1 & \vec{0} \\ \trans{\vec{0}} & N^{-1}\bdsymb{1}
\end{pmatrix} 
+ 
\begin{pmatrix}
0 &\hvec_{\perp} \\ \trans{\hvec_{\perp}} & \bdsymb{h}_{\perp}
\end{pmatrix} \right) , 
\nonumber\\
\phantom{\tframe{}}{} = 
\kappa
\begin{pmatrix}
(N^{-1}-1)\hpar & \hvec_{\perp} \\ \trans{\hvec_{\perp}} &
\bdsymb{h}_{\perp}+N^{-1}\hpar\bdsymb{1} 
\end{pmatrix} 
\in \vs{p}=\vs{s}(N),
\label{suet}\\
(\hpar,\hvec_{\perp}) \in \Rnum{N} ,\qquad
\bdsymb{h}_{\perp} \in \vs{s}(N-1) 
\nonumber
\end{gather}
and
\begin{gather}\label{suflowconx}
\tconx{}{} = \gamma_{t} \hook \conx{}{} 
= 
\begin{pmatrix}
0 & \wvec \\ -\trans{\wvec} & \bdsymb{\Theta}
\end{pmatrix} 
\in \vs{so}(N) ,\qquad
\wvec \in \Rnum{N-1} ,\qquad
\bdsymb{\Theta} \in \vs{so}(N-1) .
\end{gather}
Note the components $\hpar$, $(\hvec_{\perp},\bdsymb{h}_{\perp})$ 
correspond to decomposing 
$\tframe{} =
g(\gamma_{t},\gamma_{x})\xframe{}+(\gamma_{t})_{\perp} \hook \coframe{\perp}$ 
into tangential and normal parts relative to $\xframe{}$. 
We thus have
\begin{gather*}
 [\xframe{},\tframe{}] = 
-\kappa^2
\begin{pmatrix}
0 &\hvec_{\perp} \\ -\trans{\hvec_{\perp}} & \bdsymb{0} 
\end{pmatrix} 
\in \vs{so}(N) ,
\\ 
 [\xconx{}{},\tframe{}] = 
\kappa
\begin{pmatrix}
2\hvec_{\perp} \cdot \vvec & 
\vvec \hook \bdsymb{h}_{\perp} + \hpar \vvec \\ 
\trans{(\vvec \hook \bdsymb{h}_{\perp} + \hpar \vvec)} &
-(\vvec \otimes \hvec_{\perp} + \hvec_{\perp} \otimes \vvec)
\end{pmatrix} 
\in \vs{s}(N) ,
\\ 
 [\tconx{}{},\xframe{}] =
\kappa
\begin{pmatrix}0 & \wvec \\ \trans{\wvec} & \bdsymb{0} 
\end{pmatrix}
\in \vs{s}(N)_{\perp} .
\end{gather*}

Now the curvature equation \eqref{cartancurv} yields
\begin{gather}
\D{t}\vvec - \D{x}\wvec - \vvec \hook\bdsymb{\Theta} =
\kappa^2\hvec_{\perp} ,
\label{suveq}\\
-\D{x}\bdsymb{\Theta}+\vvec\otimes\wvec-\wvec\otimes\vvec = 
0 ,
\label{suthetaeq}
\end{gather}
which are unchanged from the case $G=SO(N+1)$ 
up to the factor in front of $\hvec_{\perp}$.  
The torsion equation \eqref{cartantors} reduces to
\begin{gather} 
0 =2\kappa^{-2} \D{x}\hpar - 2\vvec\cdot\hvec_{\perp} ,
\label{suhpareq}\\ 
0 =\wvec - \hpar\vvec - \D{x}\hvec_{\perp} 
- \vvec \hook \bdsymb{h}_{\perp} ,
\label{suweq}
\end{gather}
which are similar to those in the case $G=SO(N+1)$, plus
\begin{gather}\label{suheq}
0 = 
-\D{x}(\bdsymb{h}_{\perp}+N^{-1}\hpar \bdsymb{1})
+\vvec\otimes\hvec_{\perp}+\hvec_{\perp}\otimes\vvec .
\end{gather}

Proceeding as before, we use equations 
\eqref{suthetaeq}, \eqref{suhpareq}, \eqref{suheq} to eliminate
\begin{gather}
\bdsymb{\Theta} = 
\Dinv{x}(\vvec \otimes \wvec - \wvec \otimes \vvec) ,
\label{sutheta}\\ 
\hpar = 
\kappa^2 \Dinv{x}(\vvec\cdot \hvec_{\perp}) ,
\nonumber\\ 
\bdsymb{h}_{\perp} =
\Dinv{x}(2(1-N)^{-1} \vvec \cdot \hvec_{\perp}
\bdsymb{1}+\vvec\otimes\hvec_{\perp} + \hvec_{\perp}\otimes\vvec)\nonumber
\end{gather}
in terms of the variables $\vvec$, $\hvec_{\perp}$, $\wvec$.  
Then equation \eqref{suveq} gives a f\/low on $\vvec$,
\begin{gather*} 
\vvec_t=
\D{x}\wvec + \vvec\hook \Dinv{x}(\vvec \otimes \wvec 
- \wvec\otimes \vvec)+ \kappa^2\hvec_{\perp} 
\end{gather*}
with
\begin{gather*} 
\wvec = 
\D{x}\hvec_{\perp}+2\Dinv{x}(\vvec \cdot
\hvec_{\perp})\vvec + \vvec \hook \Dinv{x}(\vvec \otimes
\hvec_{\perp} + \hvec_{\perp} \otimes \vvec) 
\end{gather*}
obtained from equation \eqref{suweq} after we combine
$\hpar\vvec$ terms.
We thus read of\/f the operators
\begin{gather}
\Hop = 
\D{x}+\vvec \hook \Dinv{x}(\vvec \wedge\ ) ,\qquad
\Jop = 
\D{x}+2\Dinv{x}(\vvec \cdot\ )\vvec + \vvec \hook\Dinv{x}(\vvec \odot\ ) ,
\label{suHJop} 
\end{gather}
where 
$\vec{A} \wedge \vec{B} = 
\vec{A} \otimes \vec{B} - \vec{B}\otimes\vec{A}$ 
and 
$\vec{A} \odot \vec{B} = \vec{A} \otimes\vec{B}+\vec{B}\otimes\vec{A}$.

\begin{proposition}
The results in Theorem~\ref{thm1} and Corollary~\ref{cor1} 
carry over verbatim 
(with the same method of proof used in \cite{SandersWang2})
for the operators $\Hop$ and $\Jop$ here, up to a change in the scalar
curvature factor 
\begin{gather*}
\chi=-\kappa^2 = 2N/(1-N)
\end{gather*}
connected with the Riemannian geometry of $SU(N)/SO(N)$.
\footnote{ Restoring $\i$ in the identification $T_x M \simeq \i\vs{p}$
will change the sign of $\chi$. }
\end{proposition}

In particular, $\Rop=\Hop \circ \Jop$ yields a~hereditary
recursion operator
\begin{gather}
\Rop = 
\D{x}(\D{x}+2\Dinv{x}(\vvec \cdot\ )\vvec 
+ \vvec \hook\Dinv{x}(\vvec \odot\ )) 
+ \vvec \hook \Dinv{x}(\vvec \wedge
(\D{x}+\vvec \hook \Dinv{x}(\vvec \odot\ ))) 
\nonumber\\
\phantom{\Rop}{}=\D{x}^2 +2(|\vvec|^2 +(\vvec\cdot\ )\vvec) +2\Dinv{x}(\vvec \cdot\ )\vvec 
+ \vvec_x \hook\Dinv{x}(\vvec \odot\ )\nonumber\\
\phantom{\Rop=}{}
+ \vvec \hook \Dinv{x}(\vvec \wedge (\vvec \hook \Dinv{x}(\vvec \odot\ )) 
-\vvec_x \wedge\ )\label{suRop}
\end{gather}
generating a hierarchy of $O(N-1)$-invariant commuting
bi-Hamiltonian f\/lows on $\vvec$, correspon\-ding to commuting
Hamiltonian vector f\/ields
$\hvec_{\perp}^{(k)}\hook\partial/\partial\vvec$ 
and involutive covector f\/ields $\wvec^{(k)}\hook d\vvec$, 
$k=0,1,2,\ldots$ starting from
$\hvec_{\perp}^{(0)}=\vvec_x$, $\wvec^{(0)}=\vvec$.
In the terminology of \cite{AncoWolf},
$\hvec_{\perp}^{(k)}$ is said to produce
the $+(k+1)$ f\/low equation \eqref{floweq} on $\vvec$
(cf.\ Remark~\ref{rem1}).
Note these f\/lows admit the same mKdV scaling symmetry
$x\rightarrow\lambda x$, $\vvec\rightarrow\lambda^{-1}\vvec$
as in the case $SO(N+1)/SO(N)$.
They also have similar recursion relations
$\hvec^{(k)}_{\perp} = \Hop(\wvec^{(k)})$,
$\wvec^{(k+1)}=\Jop(\hvec^{(k)}_{\perp})
= \delta H^{(k+1)}/\delta \vvec$, 
$k=0,1,2,\ldots$,
in terms of Hamiltonians 
$H =H^{(k)}(\vvec,\vvec_{x},\vvec_{2x},\ldots)$. 

The $+1$ f\/low is given by $\hvec_{\perp}=\vvec_x$, yielding
\begin{gather}\label{sumkdveq}
\vvec_t = 
\vvec_{3x}+3|\vvec|^2\vvec_x+3(\vvec \cdot\vvec_x)\vvec - \chi\vvec_x .
\end{gather}
Up to the convective term, 
which can be absorbed by a Galilean transformation, 
this is a dif\/ferent vector mKdV equation compared to 
the one arising in the case $SO(N+1)/SO(N)$ for $N>2$.
The $+(k+1)$ f\/low yields a higher order version of 
this equation \eqref{sumkdveq}.

The hierarchy of f\/lows corresponds to geometrical motions of the
curve $\gamma(t,x)$ obtained in a similar fashion to the ones in the case
$SO(N+1)/SO(N)$ via identifying 
$\gamma_t \leftrightarrow\tframe{}$, 
$\gamma_x \leftrightarrow \xframe{}$,
$\covder{x}\gamma_x \leftrightarrow[\xconx{}{},\xframe{}]=\covD{x}\xframe{}$, 
and so on as before, where
$\covder{x} \leftrightarrow \covD{x}=\D{x}+[\xconx{}{},\xframe{}]$.
Note here we have
\begin{gather*}
[\xconx{}{},\xframe{}] = 
\kappa
\begin{pmatrix}
0 & \vvec \\\trans{\vvec} & \bdsymb{0} 
\end{pmatrix} ,
\qquad [\xconx{}{},[\xconx{}{},\xframe{}]] = 
2\kappa
\begin{pmatrix}
|\vvec|^2 & \vec{0} \\ \vec{0} & -\vvec \otimes \vvec 
\end{pmatrix} ,
\nonumber 
\end{gather*}
and so on. 
In addition,
\begin{gather*}
\ad([\xconx{}{},\xframe{}])\xframe{} =
\kappa^2
\begin{pmatrix}
0 & \vvec \\ -\trans{\vvec} & \bdsymb{0} 
\end{pmatrix} ,
\\ \nonumber 
\ad([\xconx{}{},\xframe{}])^2\xframe{} =
-2\kappa^3
\begin{pmatrix}|
\vvec|^2 & \vec{0} \\ \vec{0} & -\vvec \otimes \vvec
\end{pmatrix} 
= \chi [\xconx{}{},[\xconx{}{},\xframe{}]] .
\end{gather*}
Thus, for the $+1$ f\/low, 
\begin{gather*}
\hvec_{\perp}=\vvec_x ,\qquad
\hpar=
\frac{1}{2}\kappa^2 |\vvec|^2 ,\qquad
\bdsymb{h}_{\perp}=
\vvec \otimes \vvec +(1-N)^{-1} |\vvec|^2 \bdsymb{1} ,
\end{gather*} 
we obtain (through equation \eqref{suet})
\begin{gather*}
\tframe{} = 
\kappa
\begin{pmatrix}
(N^{-1}-1)\hpar & \hvec_{\perp} \\ \trans{\hvec_{\perp}} &
\bdsymb{h}_{\perp}+N^{-1}\hpar\bdsymb{1} 
\end{pmatrix} 
=\kappa
\begin{pmatrix}
-|\vvec|^2 & \vvec_x \\\trans{\vvec_x} & \vvec \otimes \vvec 
\end{pmatrix} 
\\ 
\phantom{\tframe{}}{} =
\D{x}[\xconx{}{},\xframe{}]
-\frac{1}{2}[\xconx{}{},[\xconx{}{},\xframe{}]] .
\nonumber 
\end{gather*}
Then writing these expressions in terms of $\covD{x}$ and
$\ad([\xconx{}{},\xframe{}])$, we get
\begin{gather*} 
\tframe{} = 
\covD{x}[\xconx{}{},\xframe{}] 
-\frac{3}{2}\chi^{-1} \ad([\xconx{}{},\xframe{}])^2\xframe{}
\leftrightarrow
\covder{x}^2\gamma_x -\frac{3}{2}\chi^{-1}\ad(\covder{x}\gamma_x)^2\gamma_x .
\end{gather*}
Thus, up to a sign, 
$\gamma(t,x)$ satisf\/ies a geometric nonlinear PDE given by the
non-stretching mKdV map equation \eqref{symmspmkdvmap}
on the symmetric space $SU(N)/SO(N)$.
The higher f\/lows on $\vvec$ determine higher order map equations for
$\gamma$.

The $0$ f\/low as before is $\vvec_t=\vvec_x$
arising from $\hvec_{\perp}=0,\hpar=1$,
which corresponds to the convective (traveling wave) map \eqref{convmap}.

There is also a $-1$ f\/low contained in the hierarchy,
with the property that $\hvec_{\perp}$ is annihilated by 
the symplectic operator $\Jop$
and hence lies in the kernel $\Rop(\hvec_{\perp})=0$ 
of the recursion operator. 
The geometric meaning of this f\/low is simply 
$\Jop(\hvec_{\perp})=\wvec=0$
implying $\tconx{}{}=0$ from equations
 \eqrefs{suflowconx}{sutheta}
so 
$0=[\tconx{}{},\xframe{}]=\covD{t}\xframe{}$ 
where 
$\covD{t} = \D{t}+ [\tconx{}{},\cdot]$.
Thus, as in the case $SO(N+1)/SO(N)$, 
we see from the correspondence 
$\covder{t} \leftrightarrow \covD{t}$, 
$\gamma_x\leftrightarrow \xframe{}$ 
that $\gamma(t,x)$ satisf\/ies a nonlinear geometric PDE given
by the wave map equation \eqref{wavemap} 
on the symmetric space $SU(N)/SO(N)$.

The $-1$ f\/low equation produced on $\vvec$ is again a nonlocal
evolution equation
\begin{gather}\label{suvsgflow} 
\vvec_t=-\chi \hvec_{\perp} ,\qquad
\chi=-\kappa^2 
\end{gather} 
with $\hvec_{\perp}$ satisfying
\begin{gather}\label{suwsgflow}
0 = \wvec =\D{x} \hvec_{\perp} + h\vvec + \vvec \hook
\bdsymb{h} 
\end{gather} 
where it is convenient to introduce the variables
\begin{gather*}  
\bdsymb{h} = \bdsymb{h}_{\perp}+N^{-1}\hpar\bdsymb{1} ,\qquad
h=2\kappa^{-2}\hpar=\tr\bdsymb{h}
\end{gather*}  
which satisfy
\begin{gather} 
\D{x}h = 
2\vvec \cdot\hvec_{\perp} ,
\label{suhparsgflow}\\ 
\D{x} \bdsymb{h} = 
\vvec \otimes\hvec_{\perp} + \hvec_{\perp} \otimes \vvec .
\label{suhsgflow}
\end{gather}
These equations \eqsref{suwsgflow}{suhsgflow}
determine the variables $\hvec_{\perp}$, $h$, $\bdsymb{h}$
implicitly as nonlocal functions of $\vvec$ (and its $x$ derivatives).
To proceed, we will seek an inverse local expression for $\vvec$ 
in terms of $\hvec_\perp$,
analogous to the one that arises in the case $SO(N+1)/SO(N)$. 
However, the presence of the additional variable $\bdsymb{h}$ here
leads to a quite dif\/ferent expression for the resulting f\/low on $\vvec$.
Let
\begin{gather}\label{suvsgeq}
\vvec = \alpha \D{x}\hvec_{\perp} +\beta \hvec_{\perp} 
\end{gather} 
for some expressions $\alpha(h)$, $\beta(h)$.
Substitution of $\vvec$ into equation \eqref{suhsgflow} yields
$\D{x}( \bdsymb{h} -\alpha \hvec_{\perp} \otimes\hvec_{\perp} )
= (2\beta-\D{x}\alpha) \hvec_{\perp} \otimes\hvec_{\perp}$
which is satisfied by $\beta=\frac{1}{2}\D{x}\alpha$
and
\begin{gather}\label{suhsgeq}
\bdsymb{h} = \alpha \hvec_{\perp} \otimes\hvec_{\perp} +c \bdsymb{1}
\end{gather} 
where $c$ is a constant of integration 
(and $\bdsymb{1}$ is the only available 
constant matrix that is $O(N-1)$-invariant). 
Then, substitution of $\bdsymb{h}$ and $\vvec$ 
into equation \eqref{suwsgflow} 
gives 
\begin{gather}\label{sualphaeq}
\alpha = -(h+c)^{-1} ,\qquad 
\beta = \frac{1}{2}(h+c)^{-2}\D{x}h ,\qquad 
c ={\rm const.}
\end{gather} 
which also satisfies equation \eqref{suhparsgflow}. 
Next, by taking the trace of $\bdsymb{h}$ 
from equation \eqref{suhsgeq} and using equation \eqref{sualphaeq}, 
we obtain
\begin{gather}\label{suhperpsgeq}
|\hvec_{\perp}|^2 = Nc (h+c)-(h+c)^2
\end{gather}
which enables $h$ to be expressed in terms of
$\hvec_{\perp}$ and $c$.
To determine $c$
we use the wave map conservation law \eqref{wavemapconslaw}
where, now, 
\begin{gather*}  
|\gamma_t|_{g}^{2} 
= \langle\tframe{},\tframe{}\rangle_\vs{p}
= \kappa^2( |\hvec_{\perp}|^2 +\frac{1}{2}( h^2+|\bdsymb{h}|^2) ) .
\end{gather*}
This corresponds to a conservation law admitted by
equations \eqsref{suwsgflow}{suhsgflow},
\begin{gather*}  
0 =\D{x}\left( |\hvec_{\perp}|^2
+\frac{1}{2}\big( h^2+|\bdsymb{h}|^2\big) \right) ,
\end{gather*}
and as before, 
a conformal scaling of $t$ can now be used to put
$|\gamma_t|_{g}$ equal to a constant.
A~convenient value which simplif\/ies subsequent expressions is 
$|\gamma_t|_{g}=2$, so then
\begin{gather*} 
(2/\kappa)^2 
=|\hvec_{\perp}|^2 +\frac{1}{2}\big( |\bdsymb{h}|^2+h^2 \big) .
\end{gather*}
Substitution of equations \eqsref{suhsgeq}{suhperpsgeq}
into this expression yields
\begin{gather*} 
c^2 = (2/N)^2
\end{gather*}
from which we obtain via equation \eqref{suhperpsgeq}
\begin{gather*} 
h= 2N^{-1} -1 \pm \sqrt{1-|\hvec_{\perp}|^2} ,\qquad
\alpha =|\hvec_{\perp}|^{-2}
\Big( 1 \pm \sqrt{1-|\hvec_{\perp}|^2} \Big) .
\end{gather*}
These variables then can be expressed in terms of $\vvec$ 
through the f\/low equation \eqref{suvsgflow},
namely $|\hvec_{\perp}|^2 = \chi^{-2}|\vvec_t|^2$. 
Finally, we note equations 
\eqref{suvsgeq} and \eqref{sualphaeq}
yield the explicit relation
\begin{gather}\label{suvhsgeq}
\vvec = \sqrt{\alpha} \D{x}(\sqrt{\alpha}\hvec_{\perp}) . 
\end{gather}
Hence the f\/low equation on $\vvec$ becomes
\begin{gather}\label{sunonlocalvflow}
\vvec_t=\sqrt{A_{\pm}} \Dinv{x}\big(\sqrt{A_{\pm}}\vvec\big) 
\end{gather}
where 
\begin{gather*}
A_{\pm}= 1 \pm \sqrt{1-|\vvec_t|^2} =|\vvec_t|^2 /A_{\mp} ,
\end{gather*}
with the factor $\chi$ having been absorbed by a scaling of $t$.

This nonlocal evolution equation \eqref{sunonlocalvflow} for the $-1$ f\/low
is equivalent to the vector SG equation
\begin{gather*} 
\big(\sqrt{A_{\mp}\vvec_t}\big)_x=\sqrt{A_{\pm}}\vvec 
\end{gather*}
or in hyperbolic form
\begin{gather}\label{susghyperboliceq}
\vvec_{tx}=A_{\pm}\vvec - A_{\mp}|\vvec_t|^{-2}(\vvec
\cdot \vvec_t)\vvec_t .
\end{gather}
Alternatively, through relations \eqref{suvhsgeq} and \eqref{sualphaeq}, 
$\hvec_{\perp}$ obeys a vector SG equation 
\begin{gather}\label{suSGeq}
(\sqrt{\alpha}(\sqrt{\alpha} \hvec_{\perp})_x)_t 
=\hvec_{\perp} .
\end{gather}
These vector equations \eqrefs{susghyperboliceq}{suSGeq}
possess the mKdV scaling symmetry
$x\rightarrow\lambda x$, $\vvec\rightarrow\lambda^{-1}\vvec$, 
where~$\hvec_{\perp}$ in equation \eqref{suSGeq}
has scaling weight $0$.

In \cite{AncoWolf} the symmetry-integrability classif\/ication results
show that the hyperbolic vector equation \eqref{susghyperboliceq}
admits the vector mKdV equation \eqref{sumkdveq} as a higher symmetry.
From Corollary~\ref{cor1}, it follows that 
the recursion operator~\eqref{suRop} generates a
hierarchy of vector mKdV symmetries
\begin{gather}
\vvec_{t}^{(0)} = 
\vvec_x ,
\label{su0flow}\\ 
\vvec_{t}^{(1)} =
\Rop(\vvec_x) = 
\vvec_{3x}+3(|\vvec|^2\vvec_x + (\vvec \cdot\vvec_x)\vvec) ,
\label{su1flow}\\
\vvec_{t}^{(2)} = \Rop^2(\vvec_x) =
\vvec_{5x}+5( |\vvec|^2\vvec_{3x}+3(\vvec \cdot \vvec_x)\vvec_{2x}
+ (2|\vvec_x|^2 + 3\vvec \cdot \vvec_{2x} +2|\vvec|^4)\vvec_x 
\nonumber\\ 
\phantom{\vvec_{t}^{(2)} = \Rop^2(\vvec_x) =}{}
+ (3\vvec \cdot \vvec_{3x} + 2\vvec_x\cdot \vvec_{2x} 
+ 4|\vvec|^2 \vvec \cdot \vvec_x)\vvec ) ,
 \label{su2flow}
\end{gather}
and so on, 
while the adjoint of this operator~\eqref{suRop}
generates a hierarchy of mKdV Hamiltonians
\begin{gather*} 
H^{(0)} = 
\frac{1}{2} |\vvec|^2 ,
\nonumber\\
H^{(1)} =
-\frac{1}{2}|\vvec_x|^2+\frac{1}{2}|\vvec|^4 ,
\nonumber\\ 
H^{(2)} =
-\frac{1}{2}|\vvec_{2x}|^2 -2|\vvec|^2|\vvec_x|^2 -3(\vvec \cdot
\vvec_x)^2 + |\vvec|^6 ,
\nonumber 
\end{gather*}
and so on.
All of these Hamiltonians are conserved densities for the $-1$ flow 
\begin{gather}\label{su-1flow}
\vvec_{t}^{(-1)} = \hvec_{\perp}
\end{gather}
associated to the vector SG equation \eqref{suSGeq},
and all of the mKdV symmetries commute with this flow. 
Hence these hierarchies are admitted symmetries and conserved densities
for the hyperbolic vector equation \eqref{susghyperboliceq}.
Viewed as flows,
the vector PDEs 
\eqsref{su0flow}{su2flow}, etc.,
including the $-1$ f\/low \eqref{su-1flow},
is seen to possess the mKdV scaling symmetry 
$x\rightarrow\lambda x$, $\vvec\rightarrow\lambda^{-1}\vvec$, 
with $t\rightarrow\lambda^{1+2k} t$
for $k=-1,0,1,2$, etc..
Moreover for $k\geq 0$,
all these expressions will be local polynomials in the variables
$\vvec,\vvec_x,\vvec_{xx},\ldots$
as established by results in~\cite{Sergyeyev2}
applied to the separate Hamiltonian (cosymplectic and symplectic) operators
\eqref{suHJop}\footnote{Due the doubly nonlocal form of the last term in 
the recursion operator \eqref{suRop},
the general results in \cite{Wang-thesis,Sergyeyev} are not directly
applicable.}.

\begin{theorem}\label{thm3}
In the symmetric space $SU(N)/SO(N)$
there is a hierarchy of bi-Hamiltonian flows of curves $\gamma(t,x)$
described by geometric map equations. 
The $0$ flow is a convective (traveling wave) map \eqref{convmap},
while the $+1$ flow is a non-stretching mKdV map \eqref{symmspmkdvmap}
and the $+2,\ldots$ flows are higher order analogs. 
The kernel of the recursion operator \eqref{suRop} in the hierarchy 
yields the $-1$ flow which is a non-stretching wave map \eqref{wavemap}.
\end{theorem}

\section{Concluding remarks}

In the compact Riemannian symmetric spaces $G/SO(N)$, 
as exhausted by the Lie groups $G=SO(N+1)$ and $G=SU(N)$, 
there is a hierarchy of integrable bi-Hamiltonian f\/lows of
non-stretching curves $\gamma(t,x)$,
where the $+1$ f\/low is described by 
the mKdV map equation \eqref{symmspmkdvmap}
and the $+2,\ldots$ f\/lows are higher-order analogs,
while the wave map equation \eqref{wavemap} describes a $-1$ f\/low 
that is annihilated by the recursion operator of the hierarchy.
In a parallel frame
the principal normal components along $\gamma$ for these f\/lows
respectively satisfy 
a vector mKdV equation 
and a vector hyperbolic equation, 
which are $O(N-1)$-invariant.
The hierarchies for $SO(N+1)/SO(N),SU(N)/SO(N)$
coincide in the scalar case $N=2$.
Moreover the scalar hyperbolic equation in this case is equivalent
to the SG equation.
These results account for the existence of the two known versions of
vector generalizations of the mKdV and SG equations 
\cite{AncoWolf}.

Similar results hold for hermitian symmetric spaces $G/U(N)$.
In particular,
there is a~hierarchy of f\/lows of curves in such spaces 
yielding scalar-vector generalizations of the mKdV equation
and the SG equation.
A further generalization of such results for all symmetric spaces $G/H$
will be given elsewhere \cite{forthcoming}.

\subsection*{Acknowledgments}

I am grateful to Thomas Wolf and Jing Ping Wang for stimulating discussions
in motivating this research.
I also thank the referees for many valuable comments. 
Tom Farrar is thanked for assistance with typesetting this paper.

The author acknowledges support by an N.S.E.R.C. grant.

\LastPageEnding

\end{document}